\documentclass[traditabstract]{aa}
\usepackage{graphicx}
\usepackage{natbib}
\usepackage[varg]{txfonts}
\usepackage{bm}
\usepackage{url}

\newcommand{\beq}{\begin{equation}}
\newcommand{\eeq}{\end{equation}}
\newcommand{\bea}{\begin{eqnarray}}
\newcommand{\eea}{\end{eqnarray}}
\newcommand{\req}[1]{Eq.~(\ref{#1})}

\newcommand{\Gamion}{\Gamma_\mathrm{i}}
\newcommand{\gcc}{\mbox{g~cm$^{-3}$}}
\newcommand{\kB}{k_\mathrm{B}}
\newcommand{\mel}{m_\mathrm{e}}

\newcommand{\nel}{n_\mathrm{e}}
\newcommand{\nion}{n_\mathrm{ion}}
\newcommand{\pF}{p_\mathrm{F}}
\newcommand{\TF}{T_\mathrm{F}}
\newcommand{\Tp}{T_\mathrm{pl}}
\newcommand{\xr}{x_\mathrm{r}}
\begin{document}


\title{Electron conduction opacities at the transition between moderate and strong degeneracy: Uncertainties and impact on stellar models}


   \author{Santi Cassisi\inst{1,2} 
          \and Alexander Y.\ Potekhin\inst{3}
          \and Maurizio Salaris\inst{4}
          \and Adriano Pietrinferni\inst{1}
          }

   \institute{INAF-Osservatorio Astronomico d'Abruzzo, via M. Maggini, sn.
     64100, Teramo, Italy (santi.cassisi@inaf.it)
     \and 
     INFN -  Sezione di Pisa, Largo Pontecorvo 3, 56127 Pisa, Italy
     \and Ioffe Institute, Politekhnicheskaya 26, Saint Petersburg 194021, Russia
     \and Astrophysics Research Institute,
     Liverpool John Moores University, IC2, Liverpool Science Park,
     146 Brownlow Hill, Liverpool, L3 5RF, UK 
     }

   \date{Received ; accepted }

\abstract{Electron conduction opacities are one of the main physics inputs for the calculation of low- and
 intermediate-mass stellar models, and a critical question  
is how to bridge calculations for moderate and strong degeneracy, which are necessarily 
performed adopting different methods. 
The density-temperature regime at the boundary between moderate and strong degeneracy is in fact crucial
for modelling the helium cores of red giant branch stars and the hydrogen/helium envelopes of white dwarfs. 
Prompted by recently published new, improved calculations of electron thermal conductivities and opacities
for moderate degeneracy, we study different, physically motivated
prescriptions to bridge these new computations with well
established results in the regime of strong degeneracy. We find that these different prescriptions
have a sizable impact on the predicted He-core masses at the He-flash
(up to 0.01$M_{\odot}$ for initial total masses far from the transition to non-degenerate He-cores,
and up to $\sim 0.04M_{\odot}$ for masses around the transition), 
the tip of the red giant branch (up to $\sim$0.1~mag) and the zero age horizontal
branch luminosities (up to 0.03~dex for masses far from the transition, and up to
$\sim$0.2~dex around the transition), and white dwarf cooling times (up to 40-45\% at high luminosities, and
up to $\sim$25\% at low luminosities). Current empirical constraints on the 
tip of the red giant branch and the zero age horizontal branch absolute magnitudes do not allow yet to
definitely exclude any of these alternative options for the conductive opacities. Tests against
observations of slowly-cooling faint WDs in old stellar populations will need to be performed to see whether they
can set some more stringent constraints on how to bridge calculations of conductive opacities
for moderate and strong degeneracy.
}

\titlerunning{Bridging conductive opacities for moderate and strong degeneracy}
\authorrunning{Cassisi et al. }
   \maketitle
%

\section{Introduction}\label{intro}

Calculations of the thermal conductivity of degenerate electrons and the corresponding
electron conduction opacities are a crucial input for the calculation of
stellar evolution models.
When electron degeneracy sets in the stellar interiors,
electron conduction becomes the dominant energy transport
mechanism, and the values of the electron conduction opacities are critical for the accurate
calculation of the models' thermal stratification \citep[see, e.g.,][and references therein]{cs:13}.
This is true for the interiors of brown dwarfs 
\citep[see, e.g.,][for a review]{cb:00}, the helium cores of 
low-mass stars (masses below $\sim$2.0--2.3 $M_{\odot}$) during their red giant branch (RGB) evolution
\citep[see, e.g.,][for a review of RGB models]{scw:02}, the carbon-oxygen
cores of low- and intermediate mass stars (masses below 6-7$M_{\odot}$)
during the asymptotic giant branch (AGB) phase \citep[see, e.g.,][]{cc:93}, the oxygen-neon cores
of super-AGB stars with masses between $\sim$6--7 and $\sim 10~M_{\odot}$ 
\citep[see, e.g.,][]{gbi, siess}, the cores and parts of the H-He envelopes of white dwarfs
\citep[WDs -- see, e.g.,][]{fbb},
as well as the envelopes of neutron stars \citep[see, e.g.,][for a review]{BPY}.

The calculation of the electron conduction opacity in astrophysical plasmas
is an ongoing enterprise, with a decades-long history, starting with the
works by \citet{marshak}, \citet{lee}, \citet{mestel},  elaborated in
the seminal works by \citet{SpitzerHarm}, \citet{Chapman54},
\citet{Braginskii58}, and the widely employed  calculations by
\citet{HubbardLampe69} further developed by Itoh and coworkers
\citep{FlowersItoh1, fi:79, fi:81, Itoh_84, MitakeII84, itoh:93}, and Yakovlev and coworkers \citep{YakovlevUrpin80,
UrpinYakovlev80,  RaikhYakovlev82, Yakovlev87, BaikoYakovlev95, BKPY},  
which were summarized, refined, and used in extensive calculations by
\citet[][hereafter P99]{PBHY}.

Each of  these sources of opacities had
its own limitations and shortcomings. For instance,  \citet{SpitzerHarm} considered
non-degenerate electrons, while \citet{HubbardLampe69} used different methods
of calculations in the cases of weak and strong electron degeneracy,
i.e. when $T\gg\TF$ and $T\ll\TF$, where $T$ is the temperature and
$\TF$ is the Fermi temperature (see Sect.~\ref{sect:cond1}), leaving
some gaps in the intermediate range of partially degenerate electrons,
where $T\sim\TF$. Besides, Hubbard \& Lampe tabulations covered a very
limited set  of chemical mixtures, and neither Spitzer \& H\"arm nor Hubbard \&
Lampe  took into account relativistic effects or the regime of dense matter where
the stellar plasma solidifies.  The work by Itoh's and Yakovlev's
research groups made significant improvements over the previous
results,  taking into account the effects of special relativity and more
accurate structure factors for the electron-ion plasmas, as well as the
electron-phonon scattering which replaces the electron-ion scattering in
the solid phase. Their results could be employed also to compute opacities for
arbitrary chemical mixtures, however they covered only the case of
strong electron degeneracy, i.e. a regime where $T\ll\TF$, which
is not really fulfilled in the He-cores of RGB stars or the envelopes of
WDs \citep[see, e.g.,][and the next sections for a deeper analysis of
this issue]{catelan:07}.

For the conductive opacities due to the electron-ion (\textit{ei}) scattering, a 
consistent way of filling the gap between the domains of weakly and
strongly degenerate electrons is provided by the  thermal averaging
procedure \citep[see, e.g.,][hereafter C07]{cassisi:07}, patterned after the
method previously employed by \citet{PY96} to compute finite-temperature
effects on the Shubnikov--De Haas oscillations of the electron transport
coefficients of degenerate electron-ion plasmas in quantizing magnetic
fields.\footnote{Note that here we consider non-magnetized
plasmas and focus on their thermal conductivity in the liquid phase.
For a more general
overview of the recent progress in the theory of conductivities in the
Coulomb plasmas, including the solid phase and the magnetized plasmas, see,
e.g., \citet{Potekhin15} and references therein.}
Unfortunately, this method is not applicable to the electron-electron
(\textit{ee}) scattering: To overcome this difficulty, an interpolation formula
has been proposed by C07, who have also taken into account an improved
treatment of the \textit{ee} scattering at high densities, suggested at the time
by \citet{ShterninYakovlev06}.

The electron conduction theory has undergone substantial progress in the
last decade, enabling yet refined studies of the heat  transport 
by partially degenerate electrons \citep[e.g.,][and references
therein]{Desjarlais_17,Daligault18,ShafferStarrett20}. In particular,
\citet{ShafferStarrett20} demonstrated that the \textit{ee} scattering
affects the thermal conductivity in a non-trivial way at $T\sim\TF$,
resulting in lower conductive opacities compared to the
traditional approach. This effect is especially pronounced for light
chemical elements in the regime of moderate coupling and moderate
degeneracy. Based on this theory, \citet[][hereafter B20]{Blouin_20}
calculated the conductive opacities for H and He compositions, finding
a difference by up to a factor 2.5--3  compared to C07 near the boundary
of the temperature-density domain where the new theory may be
applied. They have also shown that this decrease of the conductive
opacities has a sizable impact  on the cooling times of WD models with H
and He envelopes, such that  the age of the coolest models is reduced by
as much as $\sim$2~Gyr, compared to calculations with C07 opacities.

The important point to notice is that, as also stated by
\citet{ShafferStarrett20} and B20, the traditional (e.g., C07) results
are superior at strong degeneracy, because they, unlike B20, ensure the
known asymptotic limits at $T/\TF\ll1$. Besides, the theory underlying
the B20 results is non-relativistic and therefore it is restricted to
mass densities $\rho\lesssim10^6$ \gcc. Therefore, we need to bridge 
B20 results for mildly degenerate, non-relativistic plasmas
and the traditional opacities at higher densities. This introduces some
uncertainty, which can affect the calculation of both WD and RGB models, for
sizable portions of the helium cores of RGB models and of the H
and He envelopes of WD models, cover a range of the degeneracy
parameter  $\theta\equiv T/\TF$ that extends from a few times 0.01 to a
few times 0.10 and above.

The purpose of this paper is to investigate
different possible ways to merge B20 results at $\theta\sim1$ with the
opacities at $\theta\ll1$, studying their impact on the cooling times
of WD models, on the mass of the electron degenerate helium cores of
low-mass stellar models at the He-flash, and the resulting effect on
the RGB lifetime, the luminosities of the tip of the RGB and of the
start of quiescent core He-burning after the degeneracy has been lifted.
These luminosities are traditionally  used to constrain the distance of
old stellar populations (ages above 1--2~Gyr).

The plan of the paper is as follows. In Sect.~\ref{sect:cond} we
summarize the theoretical background to the calculations of conductive
opacities, give an overview of the recent updates for the partially
degenerate domain and discuss possible ways to treat the transition to
the strong degeneracy regime. Section~\ref{evo} presents our stellar
evolution calculations and discusses the impact of the new conductive
opacities and the related uncertainties on the results. A summary and conclusions
follow in Sect.~\ref{conclusions}.

\section{Conductive opacities}
\label{sect:cond}

In stationary and non-convective layers of a star,
the heat transport is governed by the Fourier law
$
    \bm{F} = - \lambda\,\nabla T,
$
where $\bm{F}$ is the heat flux, $T$ is the temperature, and $\lambda$
the thermal conductivity. The last quantity is related to the opacity
$\kappa$ by the equation (see, e.g., \citealt{Kippenhahn})
\begin{equation}
 \kappa =
  \frac{16\sigma T^3}{3\rho\lambda},
\label{kappa}
\end{equation}
where $\sigma$ is the Stefan-Boltzmann constant
and $\rho$ is the mass density.

In general, radiative and conductive energy transports
work in parallel, hence the total thermal conductivity is the sum
$
   \lambda=\lambda_\mathrm{r}+\lambda_\mathrm{e},
$
where $\lambda_\mathrm{r}$ and $\lambda_\mathrm{e}$  denote
the radiative (r) and electron (e) components of the thermal
conductivity $\lambda$. Accordingly,
$\kappa^{-1}=\kappa_\mathrm{r}^{-1}+\kappa_\mathrm{c}^{-1}$,
where the radiative (r) and conductive (c) opacities are related,
respectively, to $\lambda_\mathrm{r}$ and $\lambda_\mathrm{e}$ by
\req{kappa}.

The transport coefficients of the electron-ion plasmas in the case of
non-degenerate and non-relativistic electrons ($T\gg\TF$, $\xr\ll1$) and
weakly coupled ions ($\Gamion\ll1$, 
where
$
   \Gamion = ({4\pi \nion/3})^{1/3}
      {(Ze)^2/\kB T} 
      \approx
        (2.275\times10^7\mbox{~K}/T)\,
        Z^{5/3}\xr
$
is the Coulomb coupling parameter) have been calculated long ago
\citep[e.g.,][]{SpitzerHarm,Chapman54,Braginskii58} using the
classical theory by \citet{ChapmanCowling}. The theory of the thermal
conduction by electrons of arbitrary degeneracy in the fully ionized
non-relativistic stellar interior has been reviewed by
\citet{HubbardLampe69}. An extension to the degenerate electron gas with
allowance for the special relativity effects has been described in
detail by \citet{FlowersItoh1}.

\subsection{Theoretical background}
\label{sect:cond1}

When dealing with electron conduction, according to the elementary
theory based on the kinetic method and on the assumption that
the effective electron scattering rate $\nu$ does not depend on the
electron energy, $\lambda_\mathrm{e}$ can be written as \citep{Ziman}
\begin{equation}
 \lambda_\mathrm{e} = \frac32\, \frac{\nel \kB^2 T}{\mel^* \nu} 
 \mbox{~~at~}T\gg\TF,
\quad
 \lambda_\mathrm{e} = \frac{\pi^2}{3}\, \frac{\nel \kB^2 T}{\mel^* \nu} 
 \mbox{~~at~}T\ll\TF,
\label{varkappa}
\end{equation}
where $\nel$ is the electron number density, $\mel^*$ is the effective
dynamical electron mass, $\kB$ is the Boltzmann constant,  $c$ is the
speed of light, and  $\TF$ is the electron Fermi temperature.
At $T\ll \mel c^2/\kB = 5.93\times10^9$~K,
the effective electron mass is given
by $\mel^* = \mel\sqrt{1+\xr^2}$, where $\mel$ is the true electron
mass, and $\xr$ is the relativity parameter
\beq
   \xr \equiv \frac{\pF}{\mel c}
    \approx \left(\rho_6\,
    \frac{Z}{A} \right)^{1/3}.
\label{eq:xr}    
\eeq
Here, $\rho_6\equiv \rho/10^6\,\gcc$, $\pF=\hbar(3\pi^2 \nel)^{1/3}$ is
the Fermi momentum, $Z$ and $A$ are respectively the ion charge and
mass numbers. In mixtures of elements with different charge numbers $Z_j$, they
should be averaged using the number fractions $x_j=n_j/\nion$ of the
components as weights, viz.{} $\langle Z \rangle \equiv \sum_j x_j
Z_j$ (where $n_j$ is the number
of ions of species $j$ and $\nion=\nel \langle Z \rangle^{-1}$ the total number of ions,
per unit volume);  the same holds for $A_j$.  The Fermi temperature 
\beq
   \TF=\frac{\mel c^2}{\kB}\left(\sqrt{1+\xr^2}-1\right)
\eeq
 determines whether the electrons are non-degenerate ($\TF\ll T$),
strongly degenerate ($\TF \gg T$), or mildly degenerate ($\TF\sim
T$). In the non-relativistic theory, which is valid at $\xr\ll1$, we have
$\mel^*=\mel$ and $\TF \approx 3\times10^9\,(\rho_6 \,Z/A)^{2/3}$~K.

Beyond the elementary transport theory, it is still convenient to use
\req{varkappa}, but in this case $\nu$ is some \emph{effective}
collision frequency. In the fully ionized gas or liquid, the electron
heat conduction is limited by the \textit{ei} and \textit{ee}
scattering; the assumption that the scattering rates of different kinds
are mutually independent results in the so called Matthiessen's rule,
positing that the collision frequencies simply add up. Then, in the fully
ionized liquid or gas
\beq
   \nu=\nu_\mathrm{ei}+\nu_\mathrm{ee},
\label{Matt}
\eeq
where $\nu_\mathrm{ei}$ and $\nu_\mathrm{ee}$ are the frequencies of the
electron scattering on the ions and on the electrons, respectively.

The \textit{ei} collision
frequency in a strongly degenerate Coulomb liquid
 can be written in the form \citep[e.g.,][]{YakovlevUrpin80}
\begin{equation}
   \nu_\mathrm{ei} =
   4\pi Z^2 e^4 \nion\,\mel^*\,\pF^{-3}
    \Lambda(p_\mathrm{F}),
\label{tau}
\end{equation}
where $\Lambda(\pF)$ is a dimensionless Coulomb logarithm. It is
possible to compute conductivities determined by elastic \textit{ei}
scattering at arbitrary degeneracy, using a specific thermal averaging, 
which involves an energy-dependent effective collision frequency,
described by \req{tau} at every isoenergetic surface, corresponding to a
given $\pF$ \citep[e.g., C07;][]{PY96,VenturaPotekhin}.

Although the  \textit{ei} scattering is usually most important for
degenerate plasmas, the \textit{ee} scattering is non-negligible
for $Z\lesssim10$, being especially important for H and He.
\citet{Lampe68a} treated the \textit{ee} scattering using the
Chapman-Enskog solution of the quantum Lenard-Balescu kinetic equation
for the system of degenerate electrons and pointlike non-degenerate,
weakly coupled classical ions. The dynamical screening of the electrons
was treated in the random-phase approximation, applicable at $T\ll\TF$.
The author showed that the character of the scattering is different at
temperatures $T\ll\Tp$ and $\Tp\ll T\ll\TF$,
where
\beq
   \Tp = \frac{\hbar}{\kB}\,
     \left(\frac{4\pi\nel e^2 }{\mel^*} \right)^{1/2}
\eeq
is the electron plasma temperature. In a subsequent paper,
\citet{Lampe68b} applied the Chapman-Enskog solution of the quantum
Lenard-Balescu kinetic equation to the non-degenerate and weakly
degenerate electrons.

\citet{HubbardLampe69} combined these calculations with earlier results
of \citet{Hubbard66}, who considered the \textit{ei} opacities
$\kappa_\mathrm{ei}$ in a non-relativistic degenerate electron gas,
taking into account ion-ion correlations. Hubbard and Lampe provided
conductive opacities in tabular form for various chemical compositions.
Due to the use of different approximations for non-degenerate and
degenerate electrons, their tables for these two cases do not match each
other smoothly and thus contain gaps at sufficiently low temperatures
(see more details in C07).

Hubbard and Lampe used the non-relativistic theory. The expression of
$\nu_\mathrm{ee}$ for the relativistic degenerate electrons was
obtained by \citet{FlowersItoh1} at $T\ll \Tp$, and extended by \citet{UrpinYakovlev80} to 
higher temperatures, where $\Tp\lesssim T\ll \TF$.
\citet{ShterninYakovlev06} reconsidered the problem including the Landau
damping of transverse plasmons, neglected by the previous authors. They
showed that the Landau damping strongly increases $\nu_\mathrm{ee}$ in the
domain of $\xr \gtrsim 1$, $\theta\ll1$, and $T\ll \Tp$. Their fit to
to $\nu_\mathrm{ee}$ is widely used in studies of degenerate stars, and in
particular it was employed by C07.

\subsection{Suppression of opacities in partially degenerate
plasmas}
\label{sect:shaffer}

The Matthiessen's
rule (see Eq.~\ref{Matt}) results in the additivity of the \textit{ei} 
and \textit{ee} opacities
\beq
   \kappa_\mathrm{c} = \kappa_\mathrm{ei} + \kappa_\mathrm{ee}.
\eeq
and can be derived in the lowest (one-polynomial) 
approximation of the Chapman-Enskog method
\citep{Chapman54,HubbardLampe69}. Although, as stated by
\citet{HubbardLampe69}, at least the two-polynomial approximation should be
used to obtain accurate results, the accuracy provided by the Matthiessen's
rule was deemed to be sufficient for astrophysical applications 
because, using the variational principle of the kinetic theory, it can be 
shown that  $\nu_\mathrm{ei}+\nu_\mathrm{ee}\leq\nu\leq\nu_\mathrm{ei}+\nu_\mathrm{ee}+\delta\nu$,
where $\delta\nu\ll\min(\nu_\mathrm{ei},\nu_\mathrm{ee})$ (see, e.g., Chapter~7 of
\citealt{Ziman}). However, this relation implies that the shape of the
electron distribution function is the same with and without the
\textit{ee} collisions. In fact, the electron distribution
function takes on a different shape depending on whether or not the
\textit{ee} collisions occur. 

\citet{Desjarlais_17} posited a modified Matthiessen's rule in the form 
\beq
   \kappa_\mathrm{c} = S_\kappa\kappa_\mathrm{ei} + \kappa_\mathrm{ee},
\eeq
where $S_\kappa$ is a \lq{reshaping correction\rq}, representing the
indirect modification of the \textit{ei} scattering term due to the
\textit{ee} interaction. \citet{Desjarlais_17} computed
hydrogen electrical and thermal conductivities by
the QMD method using the 
Kohn-Sham density-functional theory together with a Kubo-Greenwood
response framework, and compared the results with the quantum
Lenard-Balescu solution in the regime of weak ion coupling
($\Gamion\ll1$) and  moderate degeneracy ($T\sim\TF$), where both
methods are applicable. They found that the reshaping factor
can be as low as $S_\kappa\sim0.6$. 

\citet{Daligault16,Daligault17,Daligault18} extended the formulas for
the transport coefficients of classical plasmas inside the dense plasma
regime by applying the Chapman-Enskog method to solve the quantum
Landau-Fokker-Planck (qLFP) kinetic equation. The qLFP equation extends
the classical LFP equation by accounting for the Pauli principle while
retaining the small-angle collision approximation.  This extension has
become possible due to modifications to the classical Chapman-Enskog
method. In particular, \citet{Daligault18} replaced the expansion over
the classical Sonine polynomials by a set of orthogonal \lq{quantum\rq}
polynomials. Moreover, he derived practical formulas for the calculation
of transport coefficients (electrical and thermal conductivities,
viscosity, diffusion coefficients) based on this new polynomial
expansion. He has demonstrated that with his method we  can extend the
range of validity of the classical LFP equation, determined by the
strong inequality $T\gg\TF$, to lower temperatures. For example (see
Fig.~1 in \citealt{Daligault18}, where $r_s=0.014/\xr$), the 10\%
accuracy is ensured by the classical LFP approximation at $T > 4\TF$,
while the qLFP approximation provides the same accuracy at $T>1.7\TF$
and $\xr > 0.005$, at $T>\TF$ and $\xr >0.16$, and also at
$T>0.3\TF$ and $\xr>0.25$.

In both LFP and qLFP cases the
plasma was assumed weakly coupled and non-relativistic, and these
approximations impose supplementary restrictions on the validity domain,
which can be roughly put as $\Gamion\ll1$ and $\xr \lesssim1$. In
addition, the effects of electron exchange are neglected. The exchange
effects always reduce the electron scattering rate, but no more than by
a factor of two (see \citealt{Daligault17}).

Very recently, \citet{ShafferStarrett20} have combined the qLFP theory with the concept
of mean-force scattering, where the scattering cross-sections are
calculated using the potential of mean force as the interaction
potential. This way they can account for strong coupling effects in a
plasma kinetic framework and alleviate the constraint $\Gamion\ll1$.
They found a significant suppression of the effective \textit{ee}
scattering rate in a finite temperature interval, caused by
non-monotonicities in the \textit{ee} mean-force potential as an
indirect effect of strong ion coupling. The inclusion or omission of
\textit{ee} collisions in qLFP is rather unimportant for the
electrical conductivity at low temperatures, whereas the thermal
conductivity still depends on \textit{ee} collisions at any
temperature. In the limit of a fully degenerate electron gas, $T/\TF\to0$,
the thermal conductivity obtained with this method is identical to that
of an electron gas, which is clearly unrealistic.
\citet{ShafferStarrett20} concluded that this unphysical behavior at low
temperatures is an artifact of the small-angle approximation and traced
a connection to the argument by \citet{Lampe68a}, who noted that 
large-angle \textit{ee} collisions are more strongly Pauli blocked
than \textit{ei} ones, whereas small-angle collisions are less so.
Therefore, the qLFP method, while successful over a wide range of
temperatures, still breaks down for sufficiently degenerate plasmas, in
agreement with the above-mentioned considerations by
\citet{Daligault17,Daligault18}. In addition, the method may fail in the
case of very strongly coupled Coulomb plasmas, where an accurate ion
structure factor is needed to grasp the long-range order effects
\citep{BKPY,WettaPain20}.

\subsection{Bridging the opacities of mildly and strongly 
degenerate H and He plasmas}
\label{sect:blouin}

B20 have applied the method of \citet{ShafferStarrett20} to the
calculation  of conductive opacities for pure H and He compositions. In
case of heavier elements the electron-electron interactions are less
important, so that this method is expected to produce results more
similar to the conductive opacities $\kappa_\mathrm{c}$ obtained using
the Matthiessen's rule, that is, by assuming $S_\kappa=1$. Hereafter,
following B20, we will denote the latter opacities
$\kappa_\mathrm{c}^\mathrm{Ioffe}$.  They are essentially the C07
opacities but improved as described in \citet{Potekhin15}; the
differences with the original C07 opacities for liquid  H and He plasmas
are at most within 2\%. 

B20 found a substantial reduction of the conductive opacities (corresponding
to an enhancement of the thermal conductivity) in the domain of
partial degeneracy, compared to $\kappa_\mathrm{c}^\mathrm{Ioffe}$. 
Their Tables~1 and 2 provide $\kappa_\mathrm{c}$ for pure H and pure
He compositions, which at fixed temperature $T$ reach densities
corresponding to
$\theta\equiv T/\TF$ generally between 0.2 and 0.1.
The difference with  $\kappa_\mathrm{c}^\mathrm{Ioffe}$ exceeds
a factor of 2 on the verge of this density-temperature domain.

To facilitate
the implementation of their new opacity calculations in
stellar evolution codes, B20 have devised an analytic expression for
the factor (denoted hereafter by $F$) to reduce the the traditional 
opacity to fit their numerical results. Accordingly, the 
reduced conductive opacity and enhanced thermal electron conductivity
read
 \beq
 \kappa_\mathrm{c} = \kappa_\mathrm{c}^\mathrm{Ioffe}/F,
\qquad
 \lambda_\mathrm{e}=F\lambda_\mathrm{e}^\mathrm{Ioffe}.
\label{reduct}
\eeq
The correction factor is written as
\beq
   F=1+g(x,y)H(g(x,y)),
\label{nondamp}
\eeq
 where
$x=\log(\rho/\rho_0)$, $y=\log(T/T_0)$,  $\rho_0=10^{5.45}$ \gcc{} and
$T_0=10^{8.40}$~K for hydrogen, $\rho_0=10^{6.50}$ \gcc{} and
$T_0=10^{8.57}$~K for helium, function $g(x,y)$ is a tilted scaled
Gaussian, and $H(g)$ is a correction to the Gaussian shape at large $g$
(see the explicit formulas given in B20).
This fit accurately reproduces the numerical B20 results at $\theta>0.1$
and ensures that the correction vanishes when 
$\theta\to0$. Hereafter $\kappa_\mathrm{c}^\mathrm{B20}$ and $\lambda_\mathrm{e}^\mathrm{B20}$ 
denote, respectively, the opacities and thermal conductivities given
by Eqs.~(\ref{reduct}) and (\ref{nondamp}).

\begin{figure}
\centering
\includegraphics[width=\columnwidth]{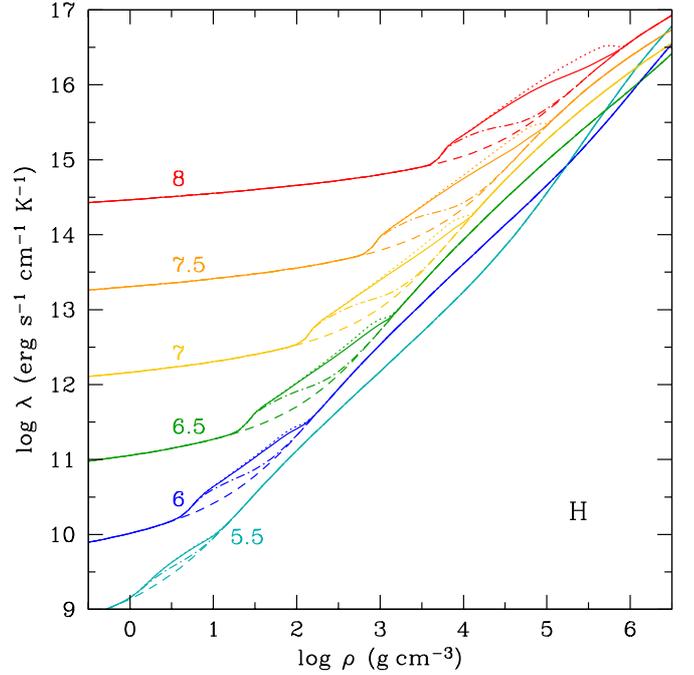}
\caption{Thermal conductivity for hydrogen as function of mass density
for several constant temperatures $T$. Dashed lines show the traditional
conductivities $\lambda_\mathrm{e}^\mathrm{Ioffe}$, and dotted lines
display the conductivities
$\lambda_\mathrm{e}^\mathrm{B20}=F\lambda_\mathrm{e}^\mathrm{Ioffe}$
enhanced by the $F$ factor in \req{nondamp}. The solid and dot-dashed
lines show the conductivities with the weakly and strongly damped
corrections given by \req{damp}, $\lambda_\mathrm{e}^\mathrm{B20wd}$
and $\lambda_\mathrm{e}^\mathrm{B20sd}$ respectively. They provide two
different transitions from the new calculations for partially degenerate
electrons $\lambda_\mathrm{e}^\mathrm{B20}$ to the traditional results
$\lambda_\mathrm{e}^\mathrm{Ioffe}$ in the strong degeneracy regime (see
text). Numbers near the curves mark $\log\,T$\,(K) values.
}
\label{fig:H}
\end{figure}

\begin{figure}
\centering
\includegraphics[width=\columnwidth]{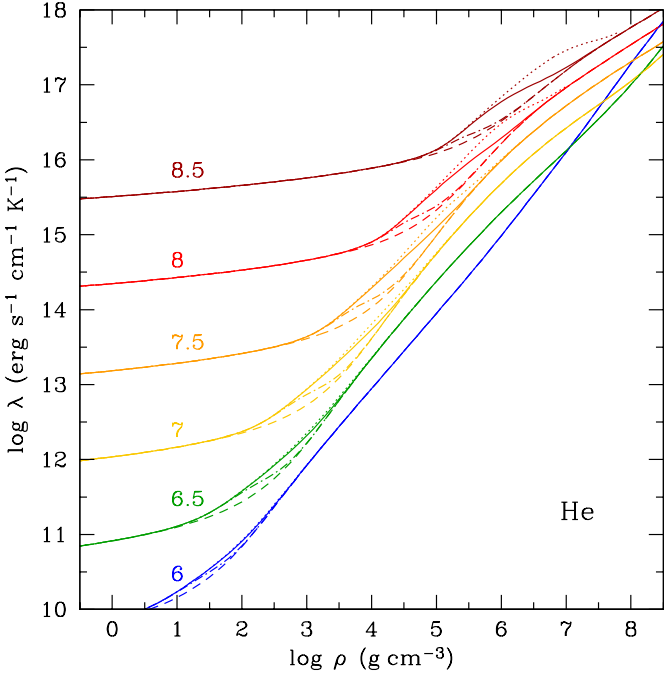}
\caption{Same as in Fig.~\ref{fig:H} but for helium.
}
\label{fig:He}
\end{figure}

The electron thermal conductivities calculated with and without the 
correction factor by B20 are shown in Fig.~\ref{fig:H} for hydrogen, and
in Fig.~\ref{fig:He} for helium, in the relevant $T$ and $\rho$ ranges. 
The convergence of $\lambda_\mathrm{e}^\mathrm{B20}$ to the traditional estimate
$\lambda_\mathrm{e}^\mathrm{Ioffe}$ 
is rather slow at high densities, if the temperature is also high.
In this case the B20 correction does not vanish until $T\ll0.1\TF$,
which is certainly far beyond the range of validity of
\citet{ShafferStarrett20} method, and most likely  overestimates the
true enhancement of the conductivities in these cases.

\begin{figure}
\centering
\includegraphics[width=\columnwidth]{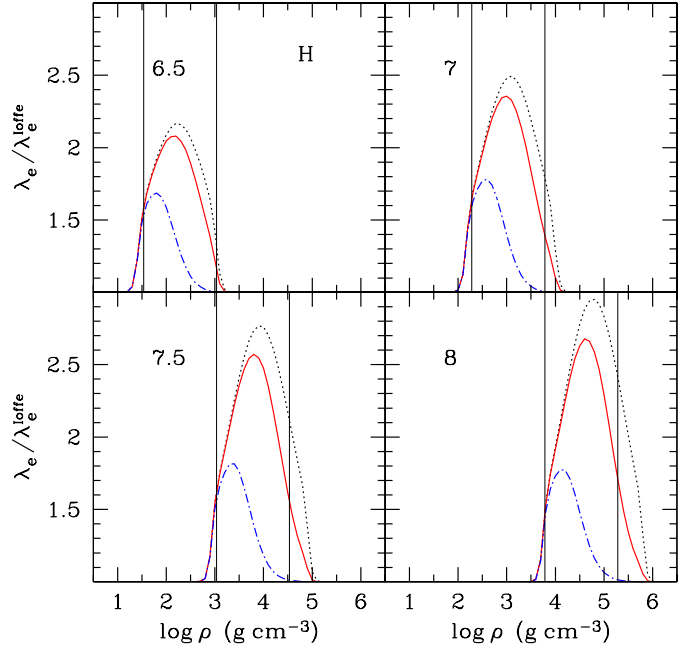}
\caption{Ratio of the electron thermal conductivity for hydrogen 
  enhanced according to B20 ($\lambda_\mathrm{e}^\mathrm{B20}$) 
  to $\lambda_\mathrm{e}^\mathrm{Ioffe}$ results, as a function of $\log\,\rho$ (dotted lines), 
  for the labelled values of $\log\,T$, with $T$ given in Kelvin. 
  Solid lines display the analogous ratio for the thermal 
  conductivities  $\lambda_\mathrm{e}^\mathrm{B20wd}$
   obtained using the weakly
  damped enhancement factor $F$.
  Dot-dashed lines show the corresponding ratio for the strong damping,
  $\lambda_\mathrm{e}^\mathrm{B20sd}$.
  The vertical lines mark the boundaries of the range of
   densities corresponding to
  $\theta=1$ and $\theta=0.1$
  (higher $\rho$ implies lower $\theta$, at constant $T$).
}
\label{figopa:H}
\end{figure}

\begin{figure}
\centering
\includegraphics[width=\columnwidth]{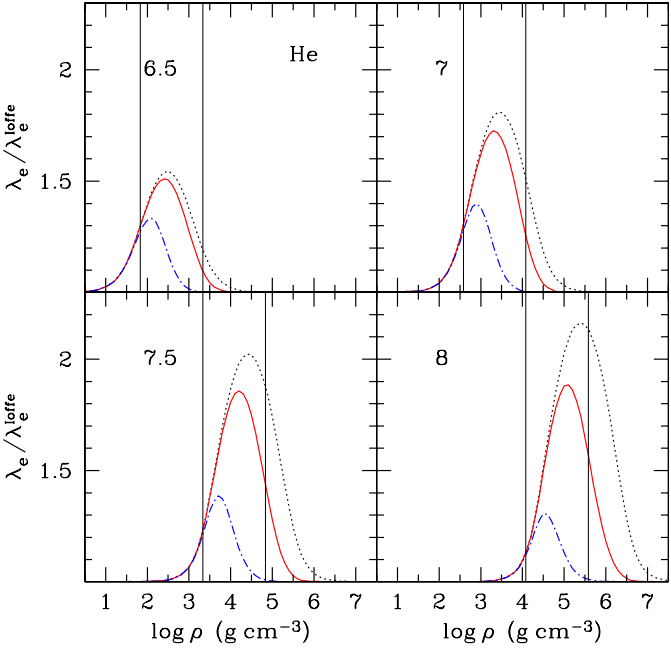}
\caption{Same as  Fig.~\ref{figopa:H} but for helium.
}
\label{figopa:He}
\end{figure}

This is shown even more clearly in Figs.~\ref{figopa:H} and ~\ref{figopa:He}, which
display the ratio $\lambda_\mathrm{e}^\mathrm{B20}/\lambda_\mathrm{e}^\mathrm{Ioffe}$ 
as a function of $\rho$ for several
temperatures $T$; densities corresponding to $\theta=0.1$ and 1.0 are also marked.
For example, at
$T>10^7$~K convergence is reached only at densities
corresponding to $\TF$ values well above $10\,T$, that is at
$\theta\ll0.1$. Besides, large differences between
the dashed and dotted lines are observed in the case of helium at
$\rho > 10^6$ \gcc, where the electrons are relativistic. 

To achieve a faster convergence to the degenerate asymptote in the
regime of strong degeneracy, we have introduced  a damping factor
$D(\theta)=(1+a\theta^{-b})^{-1}$ ($\theta\equiv T/\TF$). The damped
enhancement factor $F$ for the electron thermal conductivity (a
reduction factor for the conductive opacities) then reads
\beq
F=1+\frac{g(x,y)H(g(x,y))}{1+a\,(\TF/T)^b}.
\label{damp}
\eeq
Given that
the qLFP equation is non-relativistic, we use the non-relativistic
approximation for $\TF$ in \req{damp}. We have made two choices of the
parameters $a$ and $b$. A conservative choice (that we denote as
\lq{weak damping\rq}) is to ensure that $D(\theta)$ does not change $F$
by more than 1\% at $T>\TF$ and that it does not exceed 1\% (ensuring
that $F\approx1$) at $T<0.01\TF$. These conditions are fulfilled for 
$a=0.01$ and $b=2$. The electron conductivities obtained
using this weakly damped enhancement factor, which we denote by 
$\lambda_\mathrm{e}^\mathrm{B20wd}$ are shown in
Figs.~\ref{fig:H} and ~\ref{fig:He} 
(the corresponding opacities will be denoted by 
$\kappa_\mathrm{c}^\mathrm{B20wd}$), while 
the ratio of $\lambda_\mathrm{e}^\mathrm{B20wd}$ to $\lambda_\mathrm{e}^\mathrm{Ioffe}$
as a function of
$\rho$ is shown 
in Figs. ~\ref{figopa:H} and ~\ref{figopa:He}.

We can see that $\lambda_\mathrm{e}^\mathrm{B20}$ and 
$\lambda_\mathrm{e}^\mathrm{B20wd}$ almost coincide at $T>\TF$ (to the
left of the left vertical line in Figs. ~\ref{figopa:H} and ~\ref{figopa:He}), ensuring that our
damping does not distort the B20 results in the domain where the
underlying approximations are reliable.
The weak damping is seen to provide a good agreement with B20
results for $T$ down to $0.5\,\TF$, while at the same time it almost fully
converges to the traditional results at $T\lesssim0.03\,\TF$. 
The ratio
$\lambda_\mathrm{e}^\mathrm{B20wd}/\lambda_\mathrm{e}^\mathrm{Ioffe}$ is however 
sometimes still quite substantial (up to a factor of $\sim1.5$) at
$T\sim0.1\TF$, where $\lambda_\mathrm{e}^\mathrm{Ioffe}$ may be
already preferable, as discussed also by B20.
Indeed, as we have seen in Sect.~\ref{sect:shaffer}, results
and discussions by \citet{Daligault17,Daligault18} and
\citet{ShafferStarrett20} prompt that the approximations inherent to the
qLFP method (in particular, the small-angle scattering approximation)
may lead to an uncertainty of $\sim10\%$ at $T=\TF$ and to implausible
results at $T\ll\TF$.

We cannot therefore exclude that in reality the conductivity should 
converge to $\lambda_\mathrm{e}^\mathrm{Ioffe}$ more rapidly in the
transitional range $0.1\TF\lesssim T\lesssim\TF$.
To this purpose, we have considered a much more extreme, but probably still realistic  
\lq{strong damping\rq} choice, defined by the requirements that
$D(\theta)$ does not affect $F$ by more than 10\% at $T>\TF$ and that
$D(\theta)$ does not exceed 1\% at $T<0.1\TF$. In this case, $a=0.1$ and
$b=3$ in \req{damp}. The conductivities (see Figs.~\ref{fig:H} and~\ref{fig:He})
and opacities obtained with such
strongly damped enhancement factor will be denoted by
$\lambda_\mathrm{e}^\mathrm{B20sd}$ and 
$\kappa_\mathrm{c}^\mathrm{B20sd}$, respectively.

The ratio of $\lambda_\mathrm{e}^\mathrm{B20sd}$
to $\lambda_\mathrm{e}^\mathrm{Ioffe}$ as a function of
$\rho$ is also displayed 
in Figs.~\ref{figopa:H} and ~\ref{figopa:He}, which show how 
$\lambda_\mathrm{e}^\mathrm{B20sd}$ converges to $\lambda_\mathrm{e}^\mathrm{Ioffe}$
at $T\sim0.1\,\TF$, whilst it is almost equal to $\lambda_\mathrm{e}^\mathrm{B20}$ at $T >\TF$.
The values of $\lambda_\mathrm{e}^\mathrm{B20sd}$ may noticeably (up to
$\sim30$\%) differ from the B20 calculations already at $T\sim0.5\,\TF$, however we believe
that this strong damping option is a plausible extreme choice. As we have discussed
in Sect.~\ref{sect:shaffer}, the inaccuracy of the qLFP method may
reach 10\% at $T\sim\TF$, hence it is not unrealistic to assume still
larger inaccuracies at $T\sim0.5\,\TF$. 

\begin{figure}
\centering
\includegraphics[width=\columnwidth]{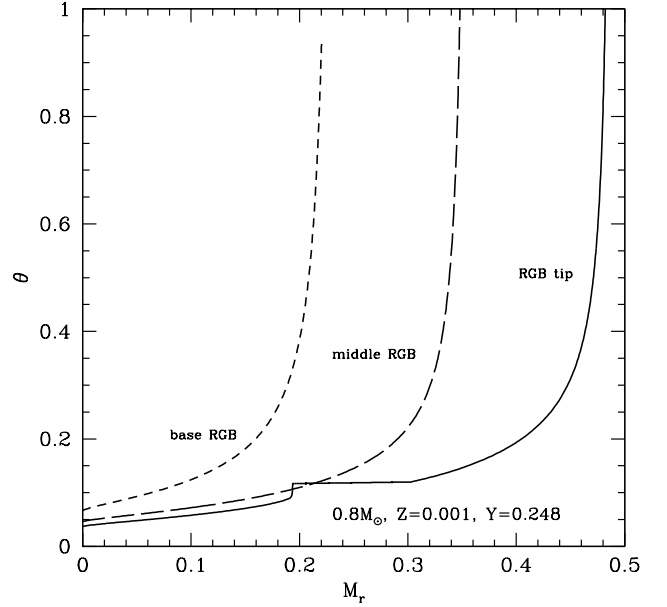}
\caption{Run of the ratio $\theta \equiv T/\TF$ as a function of the
  mass (in solar units) enclosed within a distance $r$ from the centre,
  for the He-core at three different stages of the RGB evolution of a
  model with the labelled total mass, initial helium mass fraction $Y$
  and a metal mass fraction $Z$ (metal distribution $\alpha$-enhanced,
  with [$\alpha$/Fe]=0.4, typical of Galactic halo stars) corresponding
  to [Fe/H]=$-$1.5. The discontinuity at $M_r\approx0.2M_\odot$ in the
  structure corresponding to the TRGB stellar model is due to off-center
  He-burning ignition that starts removing the electron degeneracy.}
\label{figRGBstr}
\end{figure}

\begin{figure}
\centering
\includegraphics[width=\columnwidth]{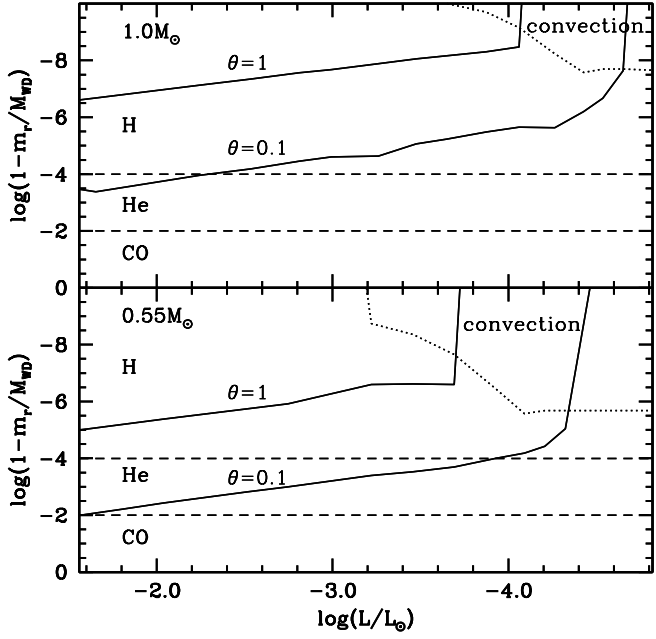}
\caption{Evolution with the surface luminosity $L$ of selected physical
  and chemical quantities across  WD models with masses equal to 0.55
  and 1$M_{\odot}$. The vertical axis displays the logarithm of the mass
  $m_r$ enclosed between the surface and a point at a distance $r$ from
  the centre, normalized to the total WD mass $M_{\mathrm{WD}}$. Dashed
  lines mark the boundaries of the carbon-oxygen core, the helium and
  the hydrogen envelope. The lower boundary of surface convection is
  denoted by dotted lines. The solid lines show the position of the mass
  layers  where $\theta$=1 and 0.1.
}
\label{figWDstr}
\end{figure}

The differences between these three choices of the electron conductivity
(B20, B20 with weak damping and B20 with strong damping) are,
by construction, maximal
around $\theta\sim0.1$ (within a factor of 3), which is a $\theta$ range
encountered in RGB He-cores and WD envelopes, as shown in
Figs.~\ref{figRGBstr} and ~\ref{figWDstr}.

Figure~\ref{figRGBstr}
displays the run of $\theta$ across the structure of the He-core  at
three selected luminosities during the RGB evolution of a typical
low-mass (0.8$M_{\odot}$), metal poor stellar model
\citep[from][]{bastiac_ae}.  In all three cases, $T/\TF$ ranges between
$\sim$0.05 at the centre of the He-core, and $\sim$1 at its outer edge.

A sketch of the internal structures of two WD models \citep[for DA WDs with He and H
  envelopes, from][]{bastiwd} and their evolution with 
the surface luminosity is shown in Fig.~\ref{figWDstr}, for masses equal to 0.55 and 1$M_{\odot}$,
bracketing the typical mass range of carbon-oxygen WDs.
For both masses, the layers where $T/\TF$ is around 0.1 are located in the He or the H layers,
depending on the model luminosity. The same is true in models with just He envelopes (for DB WDs).

\section{Effects on stellar models}
\label{evo}

In this section we quantify the effect of using
alternatively $\kappa_\mathrm{c}^\mathrm{Ioffe}$ (C07), 
$\kappa_\mathrm{c}^\mathrm{B20}$ (B20), $\kappa_\mathrm{c}^\mathrm{B20sd}$ (B20sd),
and $\kappa_\mathrm{c}^\mathrm{B20wd}$ (B20wd) opacities on RGB (and the following
horizontal branch stage) and WD models. For the RGB computations we rely
on the same stellar evolution code, physical  assumptions (including
atomic diffusion) and input physics adopted by \citet{bastiac_ae}. The
calculations by \citet{bastiac_ae} make use of C07 conductive opacities,
hence they are taken as a reference in the following discussion. For the
WD models we employ the code and physics inputs described by
\citet{bastiwd}.

\subsection{Red giant branch and horizontal branch models}\label{rgb}

We have computed models for initial masses in the range 0.8--$2.4M_\odot$ and various chemical
compositions, from the pre-main sequence stage until the He-burning ignition
at the tip of the RGB (TRGB), employing the same combinations of metallicity $Z$ and initial helium abundance
$Y$ as in \citet{bastiac_ae}.

Let us start by analysing the results for the lower-end of the explored
mass regime, i.e. for the $0.8M_\odot$ models, which are characterized
by a stronger electron degeneracy in their helium cores. The different choices of conductive
opacities have a negligible impact on the RGB lifetime but, as
expected, a sizable effect on the He-core mass at helium ignition (${\rm M_{cHe}}$).
Going from C07 to B20 opacities increases ${\rm M_{cHe}}$ by $\sim 0.01 M_{\odot}$, independent of $Z$. If the opacities
with weak damping B20wd are used instead,
${\rm M_{cHe}}$ increases by $\sim$0.007$M_{\odot}$ compared to
calculations with C07. Finally, the opacities with strong damping B20sd 
increase ${\rm M_{cHe}}$ by just $\sim$0.001$M_{\odot}$.

\begin{figure}
\centering
\includegraphics[width=\columnwidth]{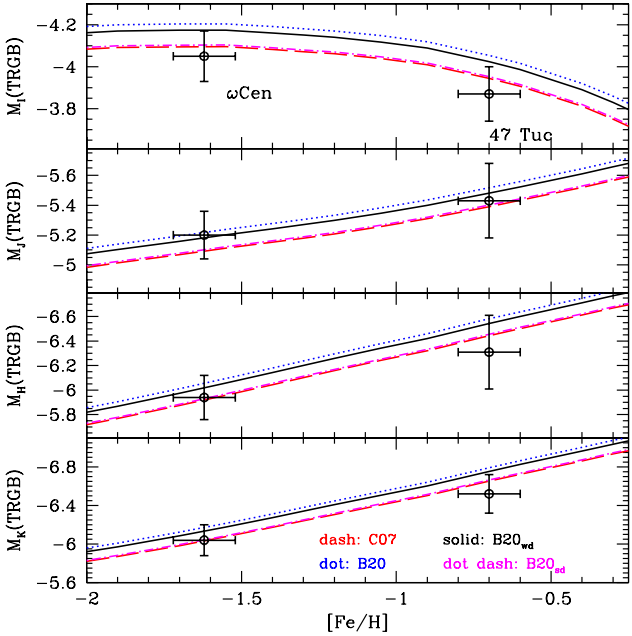}
\vskip -0.5truecm
\caption{Theoretical TRGB absolute magnitudes in the $I$ (Cousins)
  and $JHK$ (Bessel \& Brett system) filters 
  as a function of [Fe/H], for the four choices of
  conductive opacities discussed in this section (see labels and text for details).
  We also show (circles with error bars)
  the measured TRGB absolute magnitude in the Galactic globular clusters 
  47~Tuc and $\omega$~Centauri.}
\label{fig:tipobs}
\end{figure}

Given that the TRGB brightness depends on the He-core mass at the He
ignition, these differences translate to changes in the magnitude of
the TRGB, an observable quantity, also used to determine distances to old 
stellar populations in galaxies. Figure~\ref{fig:tipobs} displays the
$IJHK$ TRGB absolute magnitudes obtained from our calculations, for
models with an age of 12--13~Gyr at the TRGB, and a large range of
initial metallicities. Moving from C07 to B20 opacities makes the TRGB
brighter by about 0.1~mag in all filters (because of the larger He-core 
masses), an increase which is reduced to about 0.07~mag when
calculations with $\kappa_\mathrm{c}^\mathrm{B20wd}$ are considered
instead. The use of $\kappa_\mathrm{c}^\mathrm{B20sd}$ instead of the C07
opacities has a negligible impact of the TRGB brightness of the models.

For the sake of comparison, we also show in Fig.~\ref{fig:tipobs} the absolute magnitudes of
the TRGB determined for the Galactic globular clusters $\omega$~Centauri and
47~Tuc. We display \citet{bell:04} results, with small adjustments to take into account the recent
distance determinations by \citet{Baumg}.

The calculations using B20 and B20wd opacities predict TRGBs marginally brighter than
the observed TRGB magnitudes in the $I$ band --which incidentally have smaller
measurement errors compared to the infrared data-- when taking into account observational errors, while
in $JHK$ bands all sets of models are compatible with observations within the
error bars.
On balance these TRGB observations cannot definitely exclude any of
the displayed four choices of conductive opacities. The marginal discrepancy with
the more precise $I$-band data could for example be ascribed to some small (on the order of 0.01~mag) 
systematic errors in the calculations of the bolometric corrections,
which might affect less severely the infrared bands. 

\begin{figure}
\centering
\includegraphics[width=\columnwidth]{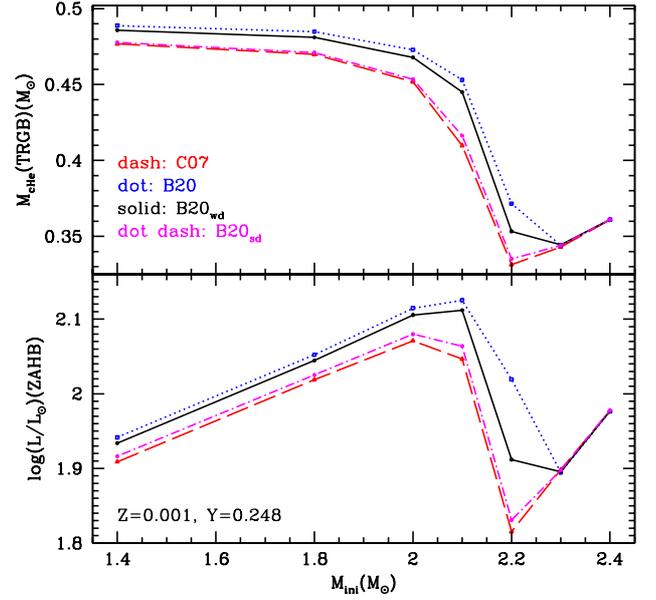}
\vskip -0.5truecm
\caption{{\it Upper panel}: Trend of the He-core mass at He-ignition as
a function of the initial total mass for models with $Z=0.001$ and
$Y=0.248$, and various assumptions about the conductive  opacities. {\it
Lower panel}: As the upper panel but for the bolometric luminosity at
the beginning of the quiescent core He-burning, after the electron
degeneracy has been lifted.} \label{fig:rgbtrans}
\end{figure}

We have also investigated the impact of these different sets of opacities on
RGB models with initial masses around the transition for the onset of electron degeneracy in the He-core.
The upper panel of Fig.~\ref{fig:rgbtrans} shows ${\rm M_{cHe}}$ at the ignition of core He-burning  
for models with $Z=0.001$ and $Y=0.248$, and masses between 1.4 and 2.4~$M_{\odot}$.
For masses larger than $\sim1.4M_\odot$, the effect of choosing a different set of opacities
increases, reaching a maximum between 2.1 and $2.2M_\odot$, to then vanish for larger masses, that 
do not develop electron degeneracy after the main sequence.
For an initial mass of 2.1--$2.2M_\odot$, the B20wd and B20 opacities increase
${\rm M_{cHe}}$ by $\sim 0.035M_\odot$ and $\sim 0.043M_\odot$, respectively.
These differences still hold at other metallicities, 
the only change being systematic shifts of the values of the initial masses of the models around the transition, 
due to the effect of the initial metallicity (and He abundance) on the mass threshold for the
onset of electron degeneracy.

\begin{figure}
\centering
\includegraphics[width=\columnwidth]{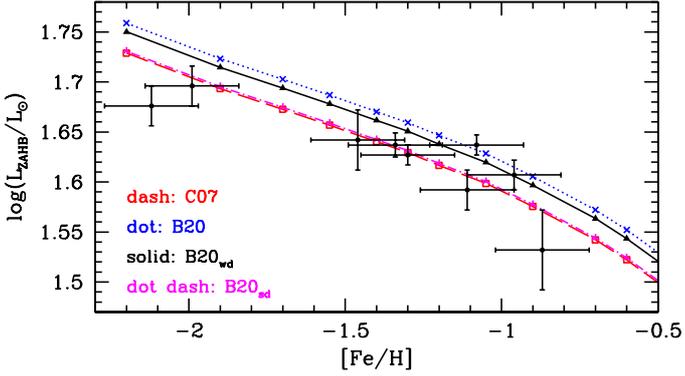}
\vskip -2.5truecm
\caption{The trend of the ZAHB brightness at the typical temperature of the RR Lyrae instability strip,
  as a function of [Fe/H] for the labelled choices of the conductive opacities. Points with error bars
  display semiempirical data for a sample of Galactic globular clusters (see text for details).}
\label{fig:zahb}
\end{figure}

These variations of the degenerate He-core masses at helium ignition
affect the properties of the following core He-burning phase, as shown
in both Figs~\ref{fig:rgbtrans} and \ref{fig:zahb}. The lower panel of
Fig.~\ref{fig:rgbtrans} shows the luminosity at the beginning of
quiescent He burning, after the degeneracy has been lifted, for the
models with initial masses between 1.4 and 2.4~$M_{\odot}$ and the
labelled initial composition  (we denote this stage by zero age
horizontal branch --ZAHB-- like for the lower mass models, which are the
theoretical counterpart of the stars that populate the horizontal
branches of globular clusters) and the labelled initial composition. The
variation of this luminosity with varying choices of the conductive
opacities mirrors qualitatively that of ${\rm M_{cHe}}$, as expected.
The effect is maximized for masses around $2.2M_{\odot}$, where
$\Delta\log(L/L_\odot)\sim0.2$~dex when passing from C07 to B20
opacities (the corresponding $\Delta\log(L/L_\odot)$ for the
1.4$M_{\odot}$ models is equal to $\sim0.03$~dex), and $\sim 0.1$ when
B20wd opacities are used instead of C07.

The change of luminosity impacts directly on the core He-burning
lifetime:  For models with $1.4\,M_\odot$ the maximum effect amounts to
a reduction  on the order of 7\% when B20 opacities are employed instead
of C07, which increases to about 18\% for models with mass around
$2.2\,M_\odot$.

Figure~\ref{fig:zahb} displays the theoretical ZAHB luminosity of lower
mass models,  at the typical temperature of the RR Lyrae instability
strip ($\log{T_{\mathrm{eff}}}=3.85$, with $T_{\mathrm{eff}}$ in K), as
a function of the initial [Fe/H], obtained using alternatively the same
four sets of conductive opacities. Again, the trends reflect the
behaviour of ${\rm M_{cHe}}$ at He-ignition. Models calculated with the
C07 opacities are the faintest ones, whilst those calculated with B20
opacities are about $\Delta\log(L/L_\odot)=0.03$ more luminous. The
calculations with the B20sd opacities are basically
identical to the C07 ones,  while the B20wd opacities
provide ZAHB models $\Delta\log(L/L_\odot)\sim 0.02$ brighter than the
C07 ones. The corresponding core He-burning lifetimes are affected at
the level of at most 6-8\% when adopting these different opacities.

The same figure displays also semiempirical ZAHB luminosities for a
sample of Galactic globular clusters, as determined by \citet{dsc} from the
pulsational properties of their RR~Lyrae stars. Due to the small sample size
  and the associated error bars, the comparison with the models does not set any
  definitive constraint on the appropriate  
  way to bridge the regimes of moderate and strong degeneracy. 
  Models based on C07 and the extreme case of the B20sd opacities, but also 
 those based on B20 and B20wd opacities are to various degrees consistent with the data.

\subsection{White dwarf models}

B20 have already shown how WD cooling models are strongly affected by
replacing $\kappa_\mathrm{c}^\mathrm{Ioffe}$ opacities in the calculations with the smaller 
$\kappa_\mathrm{c}^\mathrm{B20}$ values. Here we make similar comparisons, but
including also the cases of B20wd and B20sd opacities. We have considered a WD model with mass
$M_{\mathrm{WD}}$=1.0$M_{\odot}$, made of a carbon-oxygen core with chemical
stratification taken from the solar progenitors' models by
\citet{bastiwd}, surrounded by a helium envelope with mass equal to
$10^{-2} M_{\mathrm{WD}}$, and a more external hydrogen envelope with
mass equal to  $10^{-4} M_{\mathrm{WD}}$. The code and input physics
(except for the conductive opacities) are described in \cite{bastiwd} and
references therein. Such a high mass WD model is expected to display the
strongest sensitivity to changes of the conductive opacities, as shown 
by B20.

\begin{figure}
\centering
\includegraphics[width=\columnwidth]{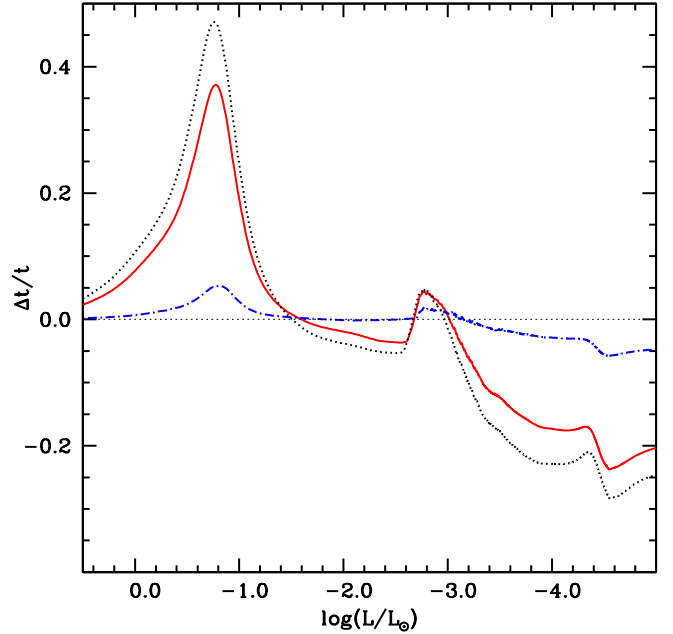}
\caption{Relative differences of the cooling times as a function of the surface luminosity, 
  among different evolutionary models of a 1.0$M_{\odot}$ DA WD.
  The dotted line displays the difference between calculations with C07 and B20 opacities (B20 cooling times
  minus C07 values at the same luminosity), the dot-dashed line the difference between B20sd and C07  
  calculations, and the solid line the difference between B20wd and C07 models.}
\label{fig:wd}
\end{figure}

Figure~\ref{fig:wd} shows the relative differences of the cooling times
as a function of the surface luminosity, among our calculations with
different opacity choices. Models calculated with B20 opacities have
longer cooling times than C07 calculations -- by up to about 40\% -- at
luminosities above $\log(L/L_\odot)\sim -1.5$, where neutrino cooling is
very efficient. As discussed by B20 \citep[see also][]{sa}, the lower
conductive opacities cause a faster cooling of the core, which reduces
the efficiency of neutrino cooling and increases the cooling times at a
given luminosity. In absolute terms, the cooling times in this phase are
relatively short, on the order of at most 100~Myr when
$\log(L/L_\odot)=-1.5$.

With decreasing luminosities, the cooling times with B20 opacities become increasingly shorter
than C07 calculations, because of the faster cooling of the structure. This trend is
temporarily broken in a narrow range of luminosities centred around
$\log(L/L_\odot)\sim -2.6$, due to the earlier start of the crystallization in the models with B20 opacities, and the
associated earlier onset of the release of latent heat and the extra energy due to carbon-oxygen phase separation
\citep[see, e.g.,][and references therein]{bastiwd}.
At the typical luminosity of the faintest observed  WDs 
($\log(L/L_\odot)\sim-4.5$) the model calculated with B20 opacities has a 
cooling age of $\sim$9.5~Gyr, about 2.5~Gyr shorter than the corresponding
calculations with C07 opacities. These differences are consistent with the results obtained by B20.

Calculations using the B20wd opacities display differences compared to
C07 models which are reduced by about 5-10\% compared to the
previous case of using B20 opacities. Like for RGB models and their core He-burning
progeny, calculations with the B20sd opacities provide results almost
identical to C07 models, with differences of the cooling times within
$\pm$5\%.

\section{Summary and conclusions}
\label{conclusions}

Electron conduction opacities are a key ingredient in the calculation of stellar models for
low- and intermediate-mass stars, and a critical issue
is how to bridge computations of conductive opacities in the regimes of
moderate ($\theta \sim 1$) and strong ($\theta \lesssim 0.1$) degeneracy, which are necessarily
calculated adopting different methods.
In fact, the density-temperature regime at the transition between moderate and strong degeneracy is crucial
for modelling the helium cores of RGB stars and the envelopes of WDs. 

We have discussed in detail the case of bridging the new, improved
conductive opacities calculated by B20 for the regime of moderate
degeneracy and the calculations by C07 in the regime of strong
degeneracy. We considered first B20 own analytical approximation,
which however converges to C07 results only at $\theta\ll$0.1, well into
the regime of strong degeneracy. We have then modified B20 formula
by introducing a physically motivated damping factor, which depends on
the ratio $\theta=T/\TF$, tuned in two alternative ways (weak and strong
damping) to converge faster than B20 fit to C07 results in the regime of
strong degeneracy. Both damping prescriptions keep almost intact
the B20 fit at $\theta > 1$. The weak damping option provides opacities still
different from C07 at $\theta=0.1$, whilst the more extreme strong
damping converges to C07 opacities at $\theta=0.1$, but changes B20
calculations already by 30\% at $\theta \sim 0.5$, in the moderate
degeneracy regime. As a consequence, these three sets of conductive
opacities have large differences (up to a factor $\sim$2) in the
critical region around $\theta\sim0.1$,  which in turn have a major
impact on the predicted RGB He-core masses (up to 0.01$M_{\odot}$ for
low-mass models far from the transition regime to non-degenerate
He-cores, and up to $\sim 0.04M_{\odot}$ for masses around the
transition), TRGB (up to $\sim$0.1~mag) and ZAHB luminosities (up to
0.03~dex for masses far from the transition, and up to $\sim$0.2~dex
around the transition), and WD cooling times (up to 40--45\% at high
luminosities, and up to $\sim$25\% at low luminosities).

Current observational constraints on TRGB and ZAHB absolute magnitudes
do not allow to categorically exclude any of these options for the
conductive opacities, also taking into account that there might be other
sources of uncertainties on the theoretical predictions for these
quantities. The much shorter cooling times predicted for faint,
slowly-evolving  WDs by calculations with both the B20 fit and the weak
damping option (compared to models calculated with opacities including
the strong damping) will need to be tested against observations of WDs
in old stellar populations.

We have updated the table of non-magnetic electron conductivities available at the Ioffe Institute
website\footnote{\url{http://www.ioffe.ru/astro/conduct/index.html}} by
implementing the correction factor in \req{damp}. We use the weak
damping as our fiducial choice by default, but we consider also the strong damping 
as a realistic extreme possibility. We have not implemented this
correction directly in the computer code presented a that website, but
provided the corresponding subroutine and envisioned a possibility of
its use (in the absence of a strong magnetic field) to correct the result of the main
computation.

\begin{acknowledgements}
  We thank our referees for constructive comments that have improved the
  presentation of our results. 
  SC acknowledges support from Premiale INAF MITiC, from INFN (Iniziativa specifica TAsP), and from
  PLATO ASI-INAF agreement n.2015-019-R.1-2018. 
The work of AYP was supported  by the Russian Science Foundation (grant 19-12-00133).

\end{acknowledgements}

\bibliographystyle{aa}
\bibliography{kcond2021}

\end{document}